# Light Propagation Prediction through Multimode Optical Fibers with a Deep Neural Network

Pengfei Fan, Liang Deng, Lei Su

School of Engineering and Materials Science, Queen Mary University of London, UK
p.fan@qmul.ac.uk, l.deng@qmul.ac.uk, l.su@qmul.ac.uk

*Abstract*—**This work demonstrates a computational method for predicting the light propagation through a single multimode fiber using a deep neural network. The experiment for gathering training and testing data is performed with a digital micro-mirror device that enables the spatial light modulation. The modulated patterns on the device and the captured intensity-only images by the camera form the aligned data pairs. This sufficiently-trained deep neural network frame has very excellent performance for directly inferring the intensity-only output delivered though a multimode fiber. The model is validated by three standards: the mean squared error (MSE), the correlation coefficient (corr) and the structural similarity index (SSIM).**

## I. Introduction

Optical multimode fibers (MMF) have recently aroused researchers' wide interest in the field of imaging, sensing and communication, due to the massive number of inside traveling modes [1-3]. While resulted from intermode dispersion and mode coupling, the light propagating inside the MMF is chaos, which causes the normally observed speckle patterns. Although scrambled to randomness, the delivery of light still emerges deterministically and linearly. So different methods have been developed to overcome this issue, such as digital optical phase conjugation [4-8], digital iterative algorithms [9-13] and other approaches based on a transmission matrix [1, 14-16]. The calibration is typically achieved by using spatial light modulators (SLMs), which can provide very high update rates and millions of degrees of freedom. The previously mentioned approaches have driven the process of turning MMFs into useful optical components, but they largely depend on technically demanding experiments and physically complicated computation.

The deep convolutional neural network (CNN), one type of deep learning techniques, has rapidly brought significant impact to various applications. In this paper, we report on an approach of predicting light propagation inside a MMF. Specifically, we demonstrate the use of a deep neural network to describe the input-output relationship though a MMF. This network uses a single binary pattern displayed on the digital micro-mirror device (DMD) as an input, and quickly outputs its corresponding intensity-only speckle image captured by the camera. The proposed CNN model has the ability to simplify such experiments and specify the intensity image for waves transmitted through a MMF. This sufficiently-trained CNN model is tested on the test dataset (not used during the training phase) with three criteria: the mean squared error (MSE), the correlation coefficient (corr) and the structural similarity index (SSIM) [17], indicating very excellent prediction performance.

## II. Experimental setup

A schematic of the experimental setup is shown in Fig.1. A DMD was used to modulate the incident laser beam with the displayed binary pattern by turning 'ON' and 'OFF' individual DMD micro-mirrors. The laser beam was subsequently coupled into the proximal end of a MMF, and the output beam from the distal end of the MMF was then recorded by the camera. (For more experimental setup details, see our last published paper [16].)

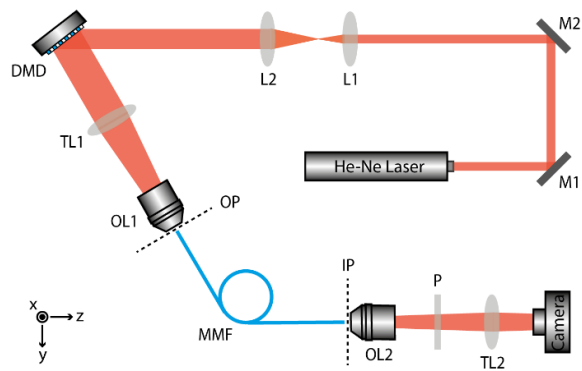

Fig. 1. A schematic of the experimental setup used to obtain the input-output data pairs for training and testing the neural network. M1, M2: Mirrors; L1, L2: Bi-Convex Lenses; DMD: Digital Micro-mirror Device; TL1, TL2: Tube Lens; OL1, OL2: Objective Lens; MMF: Multimode Fiber; P: Polarizer; OP: Object Plane; IP: Image Plane.

The input binary pattern consisting of 36×36(N = 1296) was modulated onto the DMD, and its corresponding output speckle image with 96×96(M = 9216) pixels was captured by the camera. 7780 random binary patterns at 50:50 "ON" to "OFF" ratio were generated and their corresponding output from the MMF were recorded subsequently. During the whole experiment, the system stability was monitored by calculating the correlation coefficients (above 99%) of output speckles for the same inputs over time. The patterned inputs and the captured outputs form the raw data for the training and testing stages.

III. PREDICTING LIGHT PROPAGATION WITH A DEEP NEURAL NETWORK

A. CNN Architecture

The detailed schematics of the architecture and training phase for the proposed deep neural network are depicted in Fig. 2. To infer the intricate relationship between the modulated DMD patterns and the captured MMF speckle images, we set the number of features learnt in each convolutional layer by empirically increasing the number of channels: 5, 10 and 24 respectively. The feature map dimension of the convolutional layer is empirically determined to balance the trade-off between high level of features abstraction and low network training complexity as well as short prediction output time. The size of all the kernels (filters) used throughout the network's convolutional layers is 3×3. The original input 36×36 pixels binary image is mapped into five output feature maps with 34×34 pixels by the first convolutional layer, which is followed by a batch normalization (BN) layer (which accelerates the learning convergence) [18] and a dropout layer (which prevents neural networks from overfitting) [19]. These five 34×34 pixels channels are sequentially used as input to the second convolutional layer, that reduces these images to ten 32×32 pixels output feature maps. Another three following layers are sequenced as a BN layer, a dropout layer and the third convolutional layer, which are similar with the aforementioned ones. At the output of the third convolutional layer, those ten 32×32 pixels feature maps outputted from the last layer are mapped into 24 channels with 30×30 pixels. The convolutional layers allow better representations of data with multiple levels of feature extraction, which can be combined cheaply by adding a fully-connected layer to learn a non-linear function in the feature space effectively. The flatten layer yields 24×30×30 = 21600 one-dimensional flatten units, which are subsequently activated by a ReLU activation layer. After the last BN layer and the last dropout layer, we utilised a dense layer to fully connecting the neighboring layers, which makes the model end-to-end trainable.

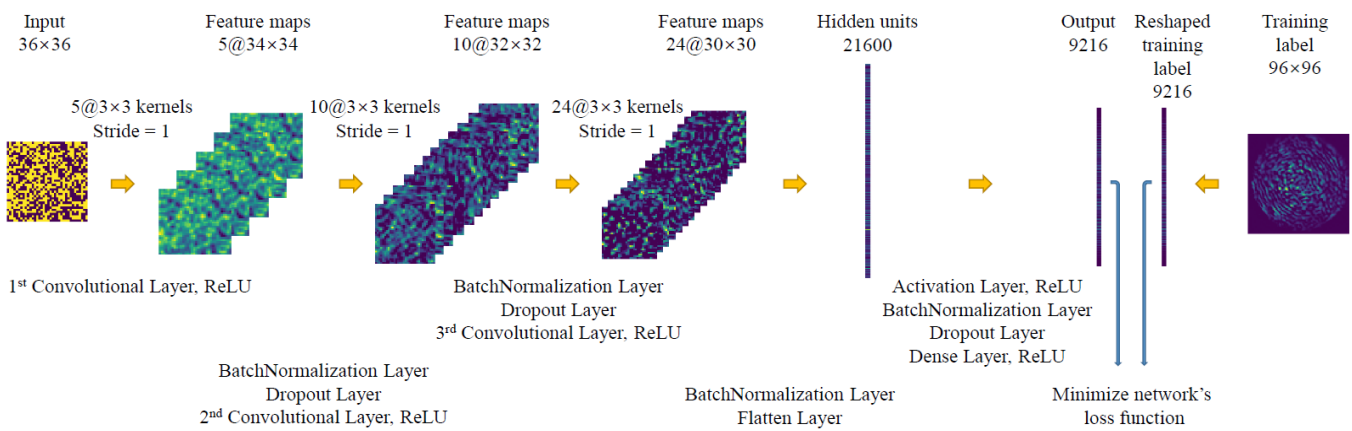

Fig. 2. Detailed architecture of the deep neural network.

B. Data pre-processing

To achieve optimal results, the network should be trained with accurately aligned input-label data pairs. The 36×36 input patterns were generated using Matlab. The 96×96 output speckle intensity-only images are acquired by camera. However, the intensity of some transmitted light would overstep the range of the camera sensor inevitably, which results in dirty data for the training and testing phase of the deep neural network. Additional data elimination is then performed on the dataset to further refine the data validity. Here, we deleted every image whose maximum intensity is under 10000 and those images with any pixel intensity greater than 65000. The thresholds for data elimination were empirically determined by the specifications of the camera (C1140-22CU, Hamamatsu) to provide the optimal balance between high learning quality and full information preservation. The intensity-only images are then bounded to the 0-255 range (implemented using *mat2gray* function provided by Matlab), which results in shorter training time and higher CNN performance. After data pre-processing, we acquired 7780 binary patterns and their corresponding images forming the input-label pairs for training, validating and testing. Out of these images, we randomly selected 7002 patches to be used as the training data set, among which 1751 pairs were used for validating the network model. To blindly quantify the average performance of the final network, the remaining 778 data pairs form our test images.

*C. CNN Training*

The total number of the network's parameters is determined by its fixed architecture. The training phase is to optimize these parameters by minimizing the network's loss function according to the training data. Table 1 shows the details of our deep neural network, presenting the arrangement of the different layers, the output shape from each layer and their maintained parameters.

TABLE I. DETAILS OF THE DEEP NEURAL NETWORK

| Layer (type) | Output Shape | Param # |
|---|---|---|
| Conv2D_1 | (34, 34, 5) | 50 |
| BN_1 | (34, 34, 5) | 20 |
| Dropout_1 | (34, 34, 5) | 0 |
| Conv2D_2 | (32, 32, 10) | 460 |
| BN_2 | (32, 32, 10) | 40 |
| Dropout_2 | (32, 32, 10) | 0 |
| Conv2D_3 | (30, 30, 24) | 2184 |
| BN_3 | (30, 30, 24) | 96 |
| Flatten_1 | 21600 | 0 |
| Activation_1 | 21600 | 0 |
| BN_4 | 21600 | 86400 |
| Dropout_3 | 21600 | 0 |
| Dense_1 | 9216 | 199074816 |

Total params: 199,164,066
Trainable params: 199,120,788
Non-trainable params: 43,278

In our implementation, a quadratic cost functions, MSE, is used to quantify the error between the predicted image and the ground truth. The MSE is defined as:

$$MSE = \frac{1}{n}\sum_{i=1}^{m}(y_i - \hat{y}_i)^2, \quad (1)$$

where $y_i$ and $\hat{y}_i$ are the expected and the predicted speckle image with $m$ pixels respectively, $n$ denotes the total amount of the training data.

The error is backpropagated through the network during the loss function minimization. Here, we use the AdaDelta optimization [20], which is a stochastic optimization method, to learn the network's parameters. The learning rate parameter is set as 0.1 and the mini-batch size is set as 32 images patches during the learning phase. The size of each filter in the convolution operation is 3×3 elements, whose entries are initialized using glorot_uniform [21], specifically,

$$w_{i,j} \sim \text{Uniform}(-\frac{\sqrt{6}}{\sqrt{n_{in}+n_{out}}}, \frac{\sqrt{6}}{\sqrt{n_{in}+n_{out}}}), \quad (2)$$

where $w_{i,j}$ is a learnable 2D kernel, $n_{in}$ and $n_{out}$ are the number of input and output channels, respectively. All the bias terms in convolutional layers are initialized with 0.

The training and validating losses are depicted in Fig. 3, specifying that the error between the network outputs and the expected labels converges steadily after 50 epochs, which indicates that the model is trained well.

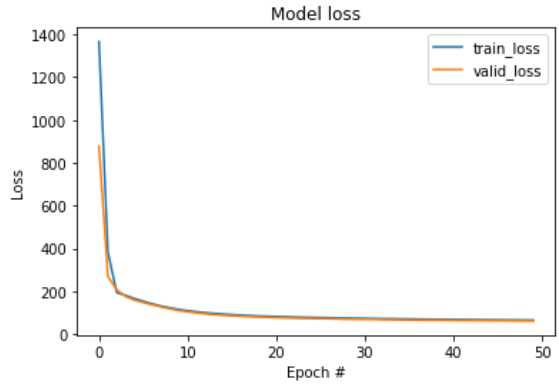

Fig. 3. Training and validation dataset errors as a function of the number of epochs.

Based on the implementation computer configuration detailed below, the Table 2 summarizes the CNN training time and other details.

TABLE II. DEEP NEURAL NETWORK TRAINING DETAILS

| Number of input-output training patches | Validation set | Training time per epoch (*sec*) | Total training time (50 epoch) |
|---|---|---|---|
| 5251 patches | 1751 images | 22 | 18.3min |

*D. CNN Testing*

A sufficiently trained network, following the predicting phase shown in Fig. 4, which receives an input of 36×36-pixel pattern and outputs a 96×96-pixel image. The output images of the CNN are plotted in Fig. 5 and Fig. 6. To numerically quantify the performance of our trained network model, we tested it using test data, as detailed in Table 3. From the perspective of regression evaluation, the MSE, the corr and the SSIM are calculated between the predicted images and the ground truth (see Fig.7 to Fig.9).

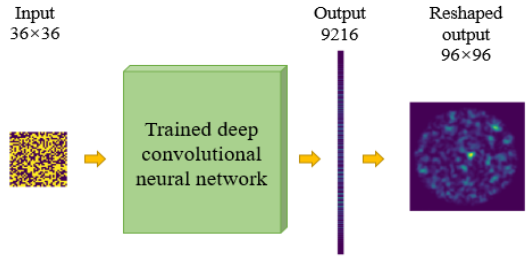

Fig. 4. The output speckle image inference phase of the deep neural network.

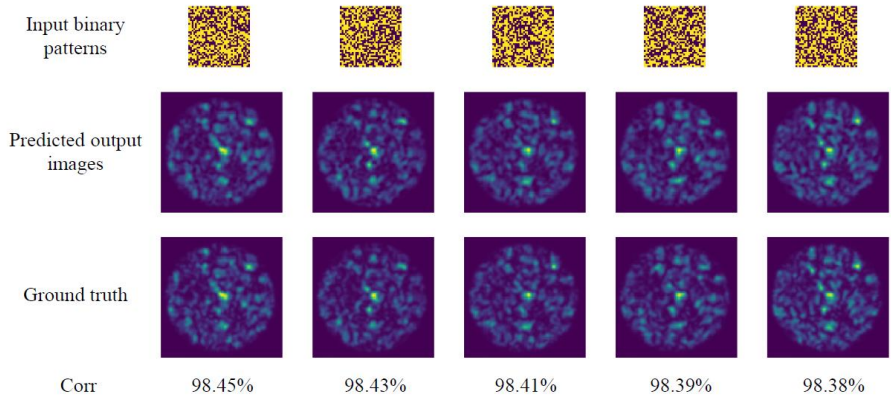

Fig. 5. The output images of the deep neural network with best performance.

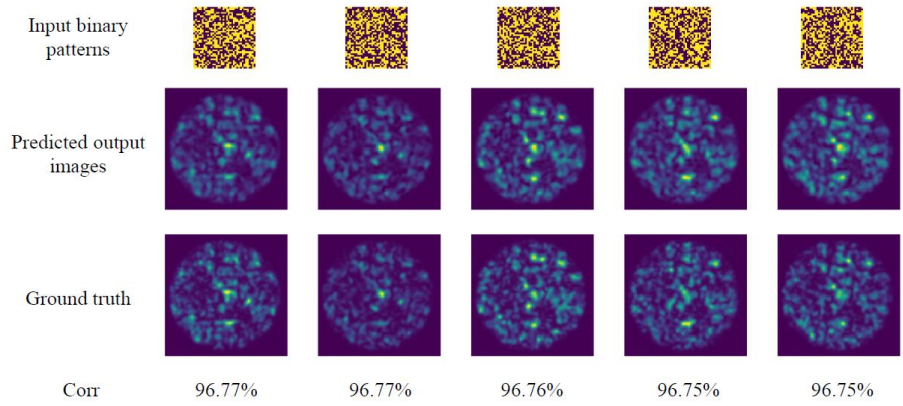

Fig. 6. The output images of the deep neural network with worst performance.

TABLE III. Average MSE, Corr, SSIM and output runtime

| Test set | MSE | Corr | SSIM | Network output runtime (*sec*) |
|---|---|---|---|---|
| 778 | 53.54 | 0.9770 | 0.9315 | 0.0224 |

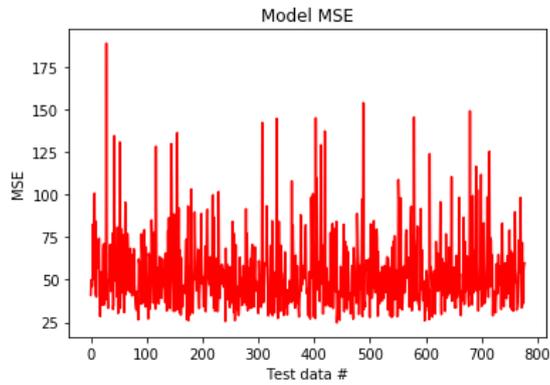

Fig. 7. Prediction MSE distribution of CNN over testing dataset.

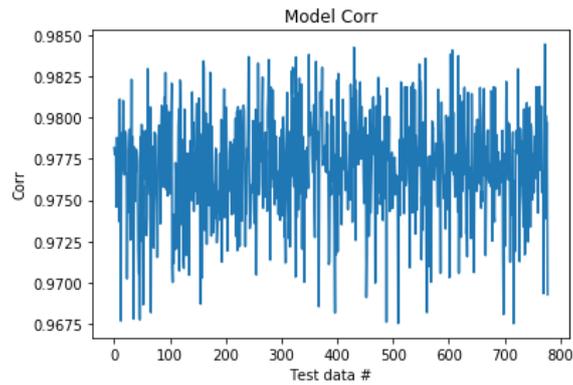

Fig. 8. Prediction Corr distribution of CNN over testing dataset.

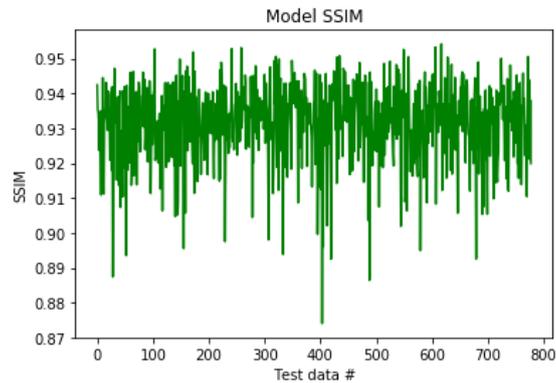

Fig. 9. Prediction SSIM distribution of CNN over testing dataset.

*E. Implementation*

The implement is conducted using Python version 3.5.2 under Keras with TensorFlow (Google) back-end. Our desktop has a GeForce GTX 1080 Ti GPU card (Nvidia), with Core i7-8700K CPU @ 3.7GHz (Intel) and 32GB of RAM, running a Windows 10 professional operating system (Microsoft). After the conclusive optimization of the network's parameters, the fixed deep neural network intakes an input stream of binary patterns each with 36×36-pixel, and outputs a 96×96-pixel image at a total time of ~ 17.5 seconds (for all the 778 images) on a laptop CPU. We used a light laptop computer with Core i7-7600K CPU @ 2.8GHz (Intel) and

8GB of RAM, running a Windows 10 professional operating system (Microsoft). The calculated runtime was averaged according to 5 different runs.

IV. CONCLUSION

In summary, it has been demonstrated that a light-weight neural network with fewer parameters allows the deep exploitation of light propagation through a MMF. The inputs and outputs are collected under a simple experimental setup (without reference beam or phase retrieval), which enables binary amplitude modulations patterned on a DMD and intensity-only images measured on a camera. The proposed deep neural network has also been examined to have a strong ability to predict the light propagation regardless of the complicated physical process in a MMF, which indicates that it is an end-to-end trainable model. We believe that the proposed model is superior to current approaches and paves the way for MMF actual applications for imaging, sensing and communication.

ACKNOWLEDGEMENTS

This work was supported by Engineering and Physical Sciences Research Council (grant number EP/L022559/1 and EP/L022559/2); Royal Society (grant numbers RG130230 and IE161214).